\documentclass{article}

\usepackage{arxiv}          
\usepackage[utf8]{inputenc} 
\usepackage[T1]{fontenc}    
\usepackage{hyperref}       
\usepackage{url}            
\usepackage{booktabs}       
\usepackage{amsfonts}       
\usepackage{nicefrac}       
\usepackage{microtype}      
\usepackage{mathrsfs}       
\usepackage{mathtools}      
\usepackage{multicol}       
\usepackage{multirow}       
\usepackage{xspace}         
\usepackage{wrapfig}        
\usepackage{subcaption}     
\usepackage{float}          
\usepackage[labelfont=bf,justification=raggedright,singlelinecheck=false]{caption}  
\usepackage{siunitx}        
\newcommand{\etal}{\emph{et~al.}\xspace} 

\title{Interpretable Machine Learning Approaches to Prediction of Chronic Homelessness}

\author{
  Blake VanBerlo\\
    VanBerlo Consulting\\
    London, Canada \\
    \texttt{blake@vanberloconsulting.com} \\
  \And
  Matthew A. S. Ross \\
   Artificial Intelligence Research and Development Lab\\
   Information and Technology Services\\
   The Corporation of the City of London\\
   London, Canada \\
   \texttt{maross@london.ca} \\
  \And
  Jonathan Rivard\\
   Homeless Prevention\\
   The Corporation of the City of London\\
    London, Canada \\
    \texttt{jrivard@london.ca} \\
  \And
  Ryan Booker\\
   Information and Technology Services\\
   The Corporation of the City of London\\
    London, Canada \\
    \texttt{rbooker@london.ca} \\
}

\begin{document}
\maketitle

\begin{abstract}
We introduce a machine learning approach to predict chronic homelessness from de-identified client shelter records drawn from a commonly used Canadian homelessness management information system. Using a $30$-day time step, a dataset for \num{6521} individuals was generated. Our model, HIFIS-RNN-MLP, incorporates both static and dynamic features of a client's history to forecast chronic homelessness $6$ months into the client's future. The training method was fine-tuned to achieve a high F1-score, giving a desired balance between high recall and precision. Mean recall and precision across $10$-fold cross validation were $0.921$ and $0.651$ respectively. An interpretability method was applied to explain individual predictions and gain insight into the overall factors contributing to chronic homelessness among the population studied. The model achieves state-of-the-art performance and improved stakeholder trust of what is usually a "black box" neural network model through interpretable AI.
\end{abstract}


\section{Introduction}
\label{sec:intro}

\subsection{Problem}
\label{subsec:problem}

Homelessness in Canada has been changing over recent years. A 2016 report claims that annually upwards of \num{235000} Canadians endure periods of homelessness, with approximately \num{35000} individuals lacking a place to stay each night~\cite{Gaetz2016}. Between 2005 and 2014, there was a downward trend in the total number of Canadians using shelters; however, the occupancy rates of shelters has been increasing~\cite{Gaetz2016}. One factor accounting for this ongoing decrease in the number of homeless individuals paired with an increase in shelter occupancy is an increase in chronic homelessness. London's Homeless Prevention division identifies an individual as chronically homelessness if they have spent $6$ or more months ($\ge \SI{180}{days}$) of the last year in a shelter, which was based on the definition of chronic homelessness outlined by the Canadian government's homelessness strategy directives~\cite{Canada2020}. In addition to this trend, the demographics of homelessness are changing in Canada. In preceding decades, older, single males are over-represented in the homeless population; in contrast, the homeless population of today is increasingly diverse, with families, women, and youth comprising a greater fraction~\cite{Gaetz2016}.

Given the diverse and evolving makeup of the modern homeless population in Canada, it would be advantageous to elucidate factors contributing to chronic homelessness to enable the predictive identification of individuals at risk of becoming chronically homeless. Shelters and municipal social services are faced with the task of preventing individuals from entering states of chronic homelessness, while acting as stewards of public resources. Machine learning models can help improve the efficiency and transparency of this process by identifying and triaging individuals at high risk of chronic homelessness. Proactive screening can inform targeted intervention before at-risk individuals suffer greater trauma and have their chronic homelessness further encumber an already overburdened shelter system ~\cite{Shinn}. This study aims to explore the efficacy of employing machine learning to predict the risk of chronic homelessness using data from the shelter system of London, Canada.

In consultation with Homeless Prevention and the London Homeless Prevention Network, it was speculated that early identification of individuals at risk of chronic homelessness may enable London’s Centralized Intake system to provide more resources to divert them from experiencing homelessness altogether. As shelters continue to adopt a housing-focused model of care, those at risk of chronic homelessness may be rapidly rehoused to further reduce over-occupancy in the shelter system. Preventive and diversionary resources are less costly overall than the reactive consumption of shelter resources by someone who has become chronically homeless. The conservation of resources via prevention of chronic homelessness would enable the shelter system to serve a greater number of individuals.

\subsection{Goals}
\label{subsec:goals}

The primary aim of this project was to develop a machine learning model to predict whether an individual would be in a state of chronic homelessness at a point $6$ months in the future. The team considered false negatives to be more harmful than false positives and therefore throughout the study, the goal was to train a proficient model that primarily minimized false negatives, while balancing a desired decrease in false positives. 

A secondary goal of the study was to gain insight into the factors that contribute to chronic homelessness in London. As a result, it was imperative that the model's predictions were interpretable. We pursued a system that would produce accurate predictions and accompanying explanations that relate client features to their predicted class. This approach of utilizing interpretable artificial intelligence (AI) has the added benefit of enabling the reduction of unintended bias and increased transparency in government-deployed automated decision systems.

A secondary goal was to release the source code and accompanying documentation under an open source license to enable other homeless services agencies across Canada to quickly train and deploy their own machine learning models for predicting chronic homelessness in their jurisdictions.  

\subsection{Precedent Research}
\label{subsec:precedent-research}

In recent years, numerous studies have applied statistical and/or machine learning concepts to model homelessness scenarios. There exist several studies and decision support tools that have been developed to serve different regions and subsets of the homeless population.

A well-known example of a decision aid that applies homelessness modelling is the Service Prioritization Decision Assistance Tool (SPDAT)~\cite{SPDAT}. With multiple versions available, the SPDAT is a screening tool that assists communities with the prioritization of homelessness prevention resources by triaging clients based on a questionnaire. As of 2015, the SPDAT was being used in communities across multiple countries, including Canada~\cite{SPDAT}. The City of London has utilized the SPDAT for over $5$ years. In essence, the SPDAT is a linear model whose features are answers to specific questions. Despite its widespread adoption, the SPDAT has its shortcomings. For instance, an American study concluded that previous versions of the Vulnerability Index-SPDAT (VI-SPDAT) struggle to hold valid and exhibit reliability, claiming that the model fell short particularly in the "Socialization and Daily Functions" and "Wellness - Health" areas~\cite{Brown2018}.

A multitude of studies have applied mathematical modelling and/or machine learning to the prediction of homelessness. A 2013 study applied Cox regression to predict whether individuals in New York City would enter shelters, and presented a screening questionnaire derived from their results ~\cite{Shinn2013}. After appropriate thresholding, the study reported an increase in recall of $26\%$ from its baseline~\cite{Shinn2013}. Also in New York, a study investigated the use of logistic regression to predict chances of readmission to shelters and length of stay~\cite{Hong2018}. These models, trained on a database of \num{6000} homeless families, attained an area under the receiver operating characteristic curve (AUC) of $0.70$ and a pseudo-R\textsuperscript{2} value of $0.069$ respectively~\cite{Hong2018}. Citizenship, age, medical history, and childhood foster care or shelter stays were reported to be the most influential features. Further, the study adapted K-means clustering to sort clients into $3$ clusters, which the authors analyzed as representative of $3$ conventionally described subtypes of homelessness: chronic, episodic, and transitional~\cite{Hong2018}. In 2016, Greer~\etal applied Cox regression to develop models predicting individual and familial entry into New York City shelters over $2$-$8$ years, which achieved AUCs of $0.90$ and $0.73$ respectively~\cite{Greer2016}.

The Economic Roundtable (ER) has undertaken several initiatives to predict homelessness. In 2011, this group developed a tool based on logistic regression that identifies the top tenth of homeless individuals ordered by expense, reporting a recall of $0.833$ and precision of $0.8$ for the task of predicting whether a person will fall in this top decile~\cite{Flaming2011}. In 2017, ER released a report on their Silicon Valley Triage Tool that predicts which homeless individuals will be the most costly users of public resources~\cite{Toros2017}. Having also investigated decision tree modelling and least-angle regression, the authors selected their logistic regression model, which achieved an AUC of $0.83$~\cite{Toros2017}. Recent work by ER focused on creating separate predictive models for identifying recently unemployed workers and for young adults at risk of becoming persistently homeless, which achieved AUCs of $0.89$ and $0.88$ respectively~\cite{Toros2019}. Both models were fitted using logistic regression and their coefficients were presented to infer the relative importance of the input features, which also exemplifies the growing importance of model interpretability~\cite{Toros2019}.

Other studies have explored different subsets of the homeless population, addressing various formulations of the problem of predicting homelessness. One study applied undisclosed predictive analytical methods to forecast first-time homelessness and return to homelessness within $12$ months, among a dataset of \SI{1.9}{million} single adults in Los Angeles County~\cite{VonWachter2019}. Chan~\etal investigated the use of logistic regression and decision trees to train interpretable models intended to function as decision aids for housing prioritization among homeless youth~\cite{Chan2017}. Finally, a 2020 Canadian study applied a custom variant of Q-learning to simulate transitions of individuals between states of homelessness, including states such as staying in a shelter, in the hospital, on the streets, and being housed~\cite{Fisher2020}. The researchers' model computes transition probability matrices on a weekly basis to generate a simulation for the population. In comparison to actual outcomes from a population dataset extending over $3$ years, the simulated population had a relative difference of $12.5\%$~\cite{Fisher2020}.

\subsection{Our Contribution}
\label{subsec:our_contribution}

The problem addressed in this paper is similar to some of the aforementioned studies. This study's focus was the development of a model that accurately predicts whether an individual will become chronically homeless in the next $6$ months and the identification of factors that influence their chronic homelessness, as well as the general drivers of chronic homelessness in London, Canada. This study is among the first to apply an artificial neural network to chronic homelessness forecasting. Further, the trained model was specifically designed to capture time-series service usage sequences in conjunction with static demographic features. Despite the inherent opaqueness of a neural network, a post-hoc interpretability method was applied that enhanced model transparency. As most precedent research pursued inherently interpretable models such as logistic regression, our results indicate that modern interpretability algorithms may be suitable to obtain stakeholders' trust in "black box" models intended to be decision aids in public services. Further, this interpretable strategy enables utilization of machine learning approaches to mathematical modelling and prediction in future homeless prevention and public service machine learning research and development. 

Perhaps of even greater importance is the replicability of our approach in other municipal jurisdictions. The source of data for our model was the City of London's Homeless Individuals and Families Information System\footnote{\url{https://www.canada.ca/en/employment-social-development/programs/homelessness/hifis.html}} (HIFIS) application which joins the service usage information for over a dozen shelters and related homeless services. The Canadian government's homelessness strategy directive mandates that municipalities are to adopt HIFIS if they lack a preexisting homelessness information management system, and are entitled to funding to assist with its deployment~\cite{Canada2020}. Care was taken to promote readability and modularity when writing our source code for all experiments to enable other HIFIS users to quickly train and deploy their own models. It is our belief that the results described in this paper could be reproduced for other jurisdictions' homeless populations given sufficient local training data.

\section{Methods}
\label{sec:methods}

\subsection{Data}
\label{subsec:data}

The raw data for this study was extracted from the database connected to the City of London's HIFIS application (version $4.0.57.30$). A SQL query was constructed to pull all records for all clients from the HIFIS database, which were subsequently saved in CSV format. The resultant raw data encapsulates interactions with social services, personal events/attributes, and demographic information. Client anonymity was preserved, as names and other identifiable information was not fetched by the query. Rather, clients were identified by a unique \textit{ClientID}. At the time of writing, London's HIFIS database contains approximately $4$ years of \num{6521} clients' records.

The model was to be trained to predict if clients were at risk of becoming or continuing to be chronically homelessness $6$ months in the future. A client was considered chronically homeless if they had at least $180$~stays over the most recent $365.25$~days. A stay was defined as $1$ or more shelter visits, occurring on the same day, that were each at least \SI{15}{minutes} in duration. Multiple visits on the same day were treated as $1$~stay. An example was therefore a $(\vec{x}, y)$ tuple, where $\vec{x}$ is a vector representing a client's state on a particular date, and $y$ is the example's corresponding ground truth. For any example, if the client met the criterion for chronic homelessness $6$~months after the example's date, then the ground truth is positive (i.e. $y=1$); otherwise, the ground truth is negative (i.e. $y=0$).

Prior to all training experiments, data was cleansed to remove any features considered to be either inconsequential or at risk of introducing unintended bias. Examples of features eliminated at the outset include height, eye colour, and hair colour. The corresponding columns in the raw data were dropped before any other pre-processing was applied.

Data preprocessing was conducted to transform the raw data to a dataset of examples which were then fed to the model. Each example included both dynamic and static features. The dynamic features were composed of numerical features describing density of usage of specific social services over the most recent $6$~time~steps. Services included features such as: number of shelter stays, number of days of case management, number of days receiving a housing subsidy, number of days in supportive housing, number of times an individual was refused service at a shelter, and number of SPDAT assessments conducted (see Appendix~\ref{apx:feature_descriptions} for a complete listing). Time steps were \SI{30}{days} long, and each dynamic feature represented the number of times a service was accessed during that time step. The time series input sequence length ($T_x$) was $6$, meaning that each dynamic service feature from the raw data resulted in $6$~features of preprocessed data -- $1$ for each of the last $6$~time~steps. In contrast, static features consisted of any feature not intended to capture recent time dependant service usage. Examples of static features include: total number of times services were accessed (since the beginning of a client's history in the HIFIS database), total monthly income, total monthly expenses, medical diagnoses, shelters they stayed at, as well as demographic information, such as age, citizenship and gender. See Appendix \ref{apx:feature_descriptions}  for a described list of all features derived from raw data.

Preprocessing of numerical features and categorical features differed in the construction of an example vector $\vec{x}$. To speed up training convergence, each numerical feature of an example was normalized by applying the operation described in Equation \ref{eqn:standard-scaling}, where $x_i$ is the $i^{th}$ feature of an example, and $\overline{x_i}$ and $\sigma_i$ are respectively the mean and standard deviation of the $i^{th}$ feature in the training set~\cite{LeCun1998}. The same transformation was applied to numerical examples in the validation and test sets, using the mean and standard deviation of the training set.
\begin{equation}
\label{eqn:standard-scaling}
x_i \leftarrow \frac{x_i - \overline{x_i}}{\sigma_i}
\end{equation}
Categorical features, were represented as one-hot encoded bit arrays. \textit{Single-valued categorical features} (SVCFs), defined as features for which an example takes on a single value (e.g. citizenship), were one-hot encoded. \textit{Multi-valued categorical features} (MVCFs) defined as features for which an example may take on any number of values were first split into a sparse array with a new feature for every possible value of the original MVCF.  For instance, health issues are a MVCF because a client may have $0$ or more health issues. MVCFs were transformed into sparse bit arrays, where each element was a Boolean flag indicating the presence or lack thereof of each value in the MVCF's domain. For example, the feature \textit{IncomeType} is split into the following binary features: \textit{IncomeType\_Pension}, \textit{IncomeType\_Student Loans(s)}, \textit{IncomeType\_Old Age Security}, etc. 

In cases of missing data, assumptions were made to impute the blank fields. If a client did not have service records during a particular time step, their service usage was set to $0$. Any other numerical features were also set to $0$ if the client was missing records for that feature. Exceptions to this rule were made for client weight (\textit{ClientWeightKG}) and recent SPDAT score (\textit{TotalScore}), which were set to $-1$ if their values were nonexistent. Any absent values for SVCFs were set to "Unknown". If a record had no values for a MVCF, the binary features corresponding to each possible value were set to $0$.

\begin{figure}[h!]
    \centering
    \includegraphics[width=\linewidth]{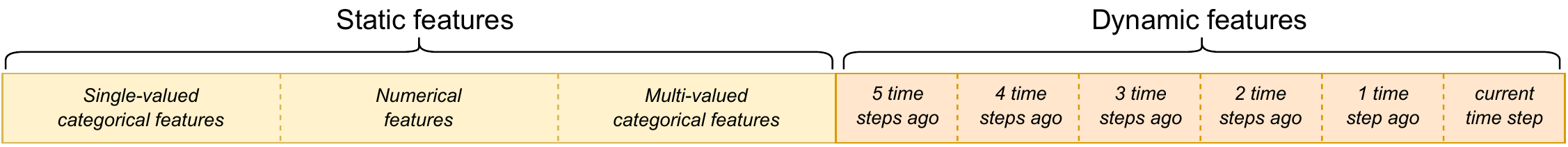}
    \caption{A breakdown of the composition of a feature vector ($\vec{x}$) for an example in the preprocessed dataset.}
    \label{fig:preprocessed-example} 
\end{figure}

\begin{figure}
    \centering
    \includegraphics[width=\linewidth]{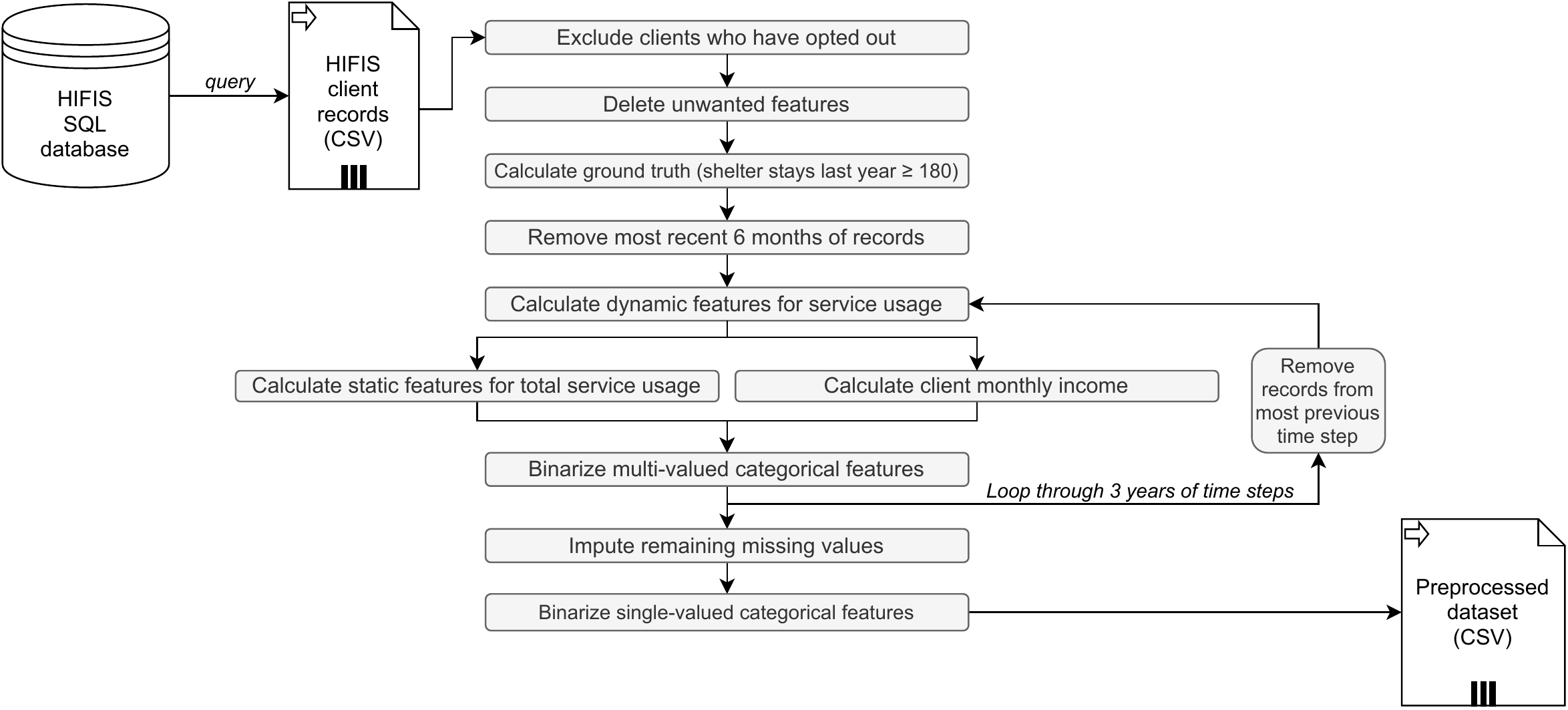}
    \caption{Major steps in data preprocessing, describing how records from the HIFIS database are transformed into the preprocessed dataset.}
    \label{fig:preprocessing-flowchart} 
\end{figure}

See Figure \ref{fig:preprocessed-example} for a visual breakdown of an example that has been preprocessed. The  preprocessed dataset was a table, indexed by \textit{ClientID} and \textit{Date}, where \textit{Date} is the final date of the current time step. The dataset contained records for each client dating back to their first records of service in the HIFIS database. Figure \ref{fig:preprocessing-flowchart} summarizes the entire preprocessing procedure that transforms raw client records into $(\vec{x},y)$ examples. At the time of writing, the preprocessed dataset contained \num{109575} examples, $6.56\%$ of which had a positive ground truth.

Prior to training, the data was partitioned into training, validation and test sets. As is customary for time series scenarios, validation data comprised the end segment of the dataset~\cite{Bergmeir2012}. In keeping with this paradigm, the validation and test sets were taken to be the second-most and most recent partitions of data respectively, where each such partition was composed of all clients' preprocessed records from $1$-$2$ time steps.

\subsection{Model}
\label{subsec:model}
A neural network model was designed to capture any time dependencies between dynamic features, in combination with the information contained in the static features. Our chosen model architecture, dubbed \textit{HIFIS-RNN-MLP}, consisted of $2$ components: a recurrent neural network (RNN) and a multilayer perceptron (MLP). Inspiration for this model arose from Hsu~\etal's application of a hybrid model that combined a RNN and random forest (RF) components~\cite{Hsu2019}. To predict credit card defaults. This approach to machine learning based risk assessment can be applied to the prediction of chronic homelessness as both problems involve using dynamic and static features to predict rare undesirable events. The choice to employ a MLP as the second component made our entire model architecture differentiable and thus end-to-end trainable. 

\begin{figure}
    \centering
    \includegraphics[width=0.6\linewidth]{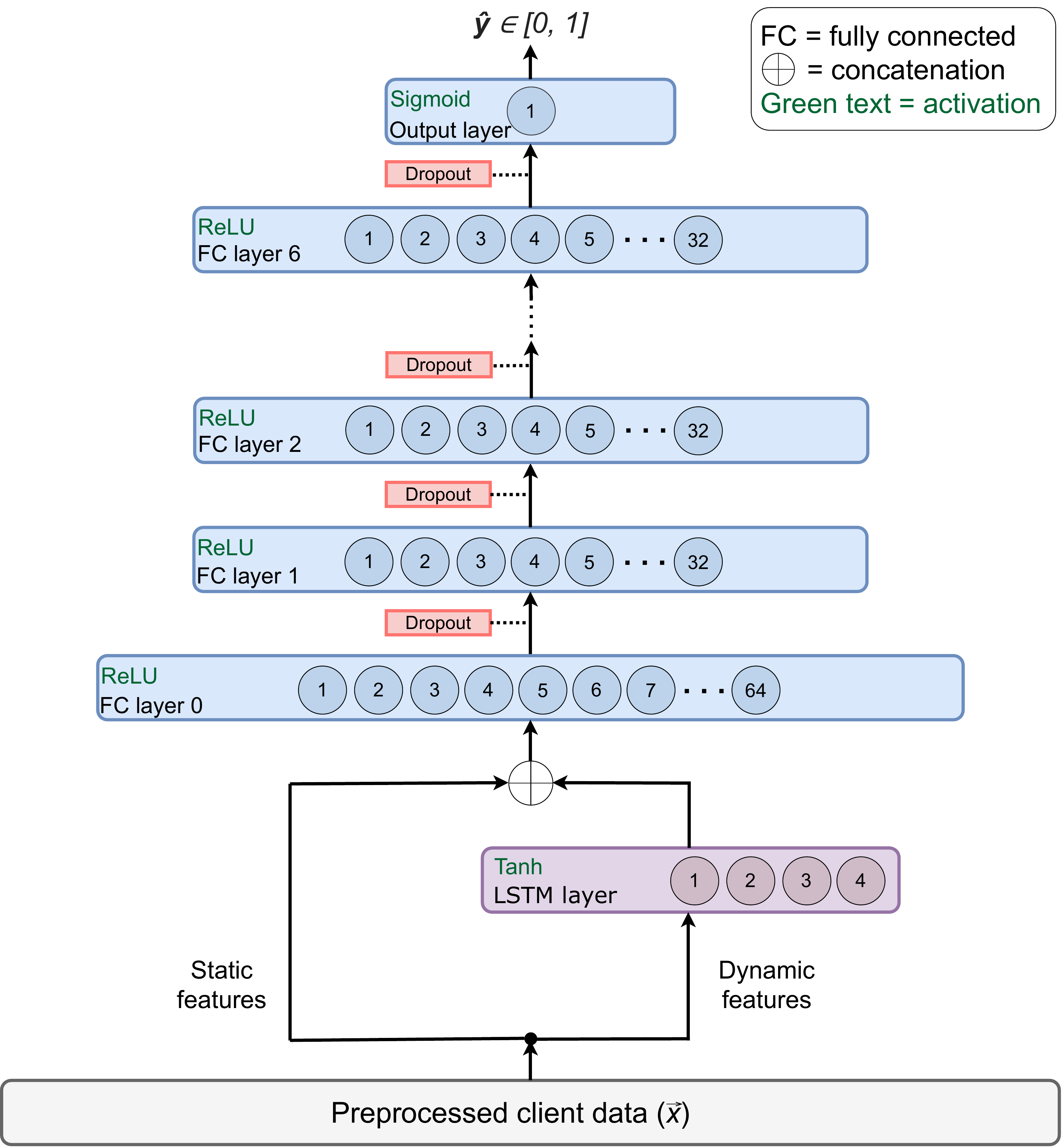}
    \caption{The HIFIS-RNN-MLP architecture. Dynamic features are passed to an LSTM layer before being concatenated with the static features to be fed to a series of fully connected layers.}
    \label{fig:nn-architecture} 
\end{figure}

Examples were fed into our model as feature vectors. The dynamic features in the example were isolated and reshaped into a matrix, then passed to the RNN that consisted of long short-term memory (LSTM) cells. The output of the LSTM and its hidden state outputs for each time step were concatenated with the static features, then passed to the MLP. The final layer of the MLP was a single neuron with sigmoid activation, whose output represented the model's assignment of probability that the client would be chronically homeless $6$ months after the date of the example. The classification threshold was set to $0.5$. The output neuron's bias was initialized to the natural logarithm of the ratio of positive to negative ground truth examples in the training set. When training neural network classifiers on imbalanced data with few positives, this initialization technique accelerates convergence by coercing the model to naively predict a low probability at the start of training~\cite{Karpathy2019}. The model's architecture is portrayed in Figure \ref{fig:nn-architecture}.

A selection of regularization methods were applied to combat overfitting. First, the L2 regularization penalty (${\gamma = \num{1.78e-3}}$) was applied to all pre-output fully connected layers in the MLP component. Additionally, dropout was applied to all fully connected layers in the MLP component at a rate of $0.44$~\cite{Srivastava2014}. Lastly, early stopping was employed to halt training once validation loss did not decrease for $15$ epochs~\cite{Prechelt1998}, and the model weights were frozen at the epoch corresponding to the minimum validation loss.

The model was trained using the Adam optimization method~\cite{Kingma2014} at a learning rate ($\alpha$) of \num{1e-3} for a maximum of $300$ epochs. Early stopping typically discontinued the training loop prior to reaching $300$ epochs. Equation \ref{eqn:custom-f1-loss} shows the custom weighted F1 loss function employed during training.
\begin{equation}
\label{eqn:custom-f1-loss}
\mathscr{L} \vcentcolon= 1 - {\textit{F1}}_{weighted} = 1 - \frac{2PR}{(2 - \frac{2}{w_r + 1})P + (\frac{2}{w_r + 1})R}
\end{equation}
\begin{multicols}{2}
  \begin{equation}
  \label{eqn:precision}
    P = \frac{\textit{true positives}}{\textit{predicted positives}} = \frac{\sum y\cdot\hat{y}}{\sum \hat{y}}
  \end{equation}
  \begin{equation}
  \label{eqn:recall}
    R = \frac{\textit{true positives}}{\textit{actual positives}} = \frac{\sum y\cdot\hat{y}}{\sum y}
  \end{equation}
\end{multicols}

\begin{figure}
    \begin{center}
        \includegraphics[width=0.6\textwidth]{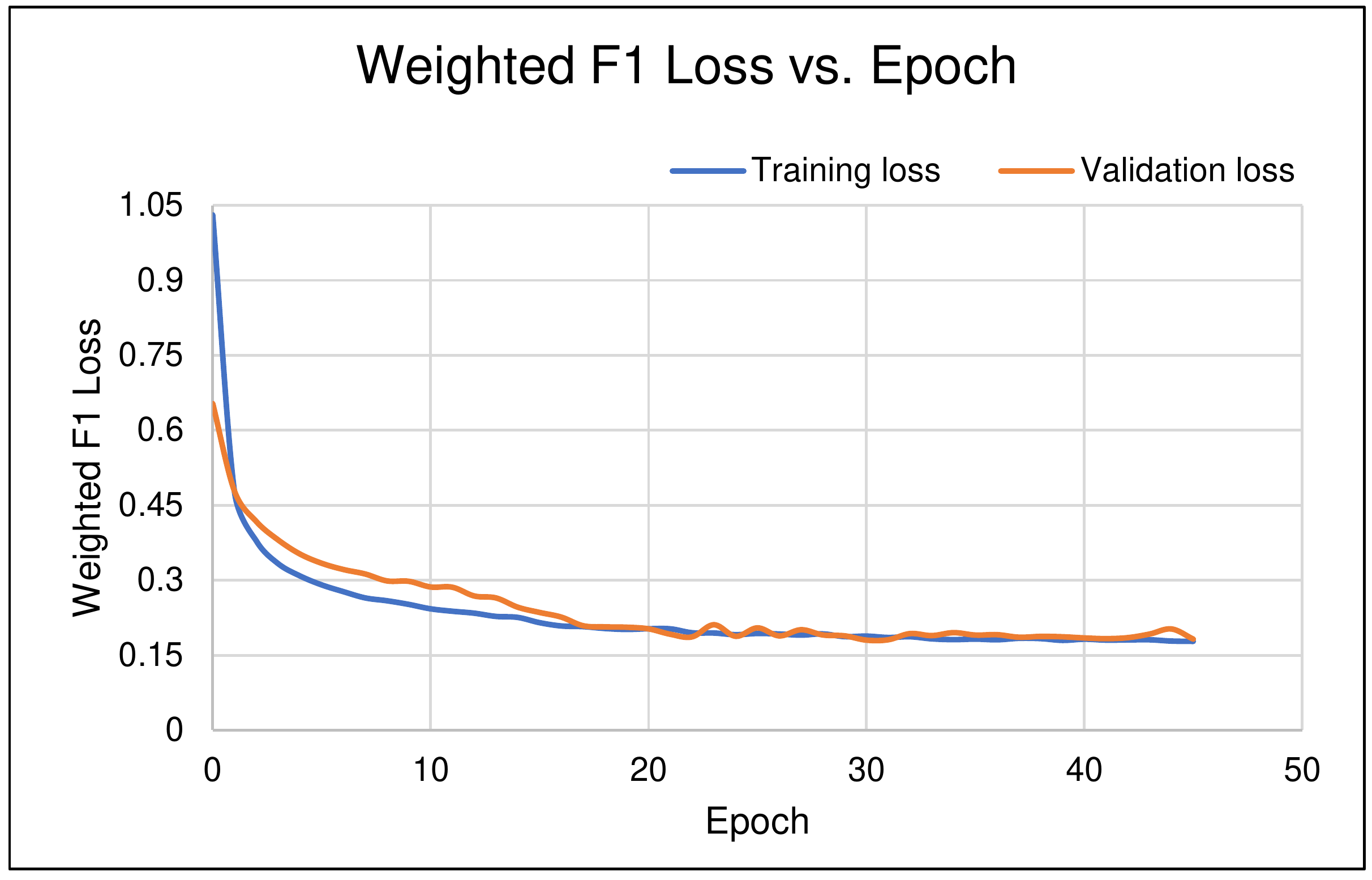}
    \end{center}
    \caption{A sample of the training curves for the HIFIS-RNN-MLP model demonstrating convergence with the weighted F1 loss function.}
    \label{fig:training-curve} 
    \vspace{-10pt}
\end{figure}

In Equation \ref{eqn:custom-f1-loss}, $P \in [0,1]$ is precision, $R \in [0,1]$ is recall, and $w_R \in \mathbb{R}^+$ is the recall weight. In Equations \ref{eqn:precision} and \ref{eqn:recall}, model predictions and ground truths are represented by $\hat{y} \in [0,1]$ and $y \in \{0,1\}$ respectively. Note that precision and recall are computed here using the probabilistic predictions $\hat{y}$, guaranteeing differentiability. Setting $w_R > 1$ more harshly penalizes the model if recall is low. In training our model, we found that $w_R = 4.5$ achieved a desirable balance between precision and recall that favoured the latter. See Figure \ref{fig:training-curve} for an example of training and validation curves using the weighted F1 loss function.

A handful of hyperparameters were studied to optimize model performance on validation and test sets. Bergstra and Bengio demonstrated that randomly searching the hyperparameter space is equally as effective and less computationally burdensome than grid search~\cite{Bergstra2012}. Several of these random hyperparameter searches were completed, narrowing down optimal ranges of the hyperparameters after each experiment. Table \ref{tab:hyperparameters} lists the final hyperparameter values adopted for the HIFIS-RNN-MLP model.

\begin{wraptable}{R}{0.5\linewidth}
    \centering
    \vspace{-20pt}
    \scalebox{0.82}{
    \begin{tabular}{c c}
        \toprule
        Hyperparameter          & Value \\
        \midrule
        \# LSTM units in RNN                                & \num{4} \\
        \# Fully connected layers in MLP                    & \num{6} \\
        \# Nodes in first fully connected layer of MLP      & \num{64} \\
        \# Nodes in remaining fully connected layers of MLP & \num{32} \\
        Dropout rate                                        & \num{0.44} \\
        L2 regularization parameter ($\gamma$)              & \num{1.78e-3} \\
        Learning rate ($\alpha$)                            & \num{1e-3} \\
        Batch size                                          & \num{1024} \\
        \bottomrule
    \end{tabular}}
    \caption{Final hyperparameter values}
    \vspace{-5pt}
    \label{tab:hyperparameters}
\end{wraptable}

The code used to arrive at the results presented in this paper was written in Python~$3$ and is publicly accessible via our GitHub repository \footnote{\url{https://github.com/aildnont/HIFIS-model}}. All training experiments were conducted on a computer running Windows~10, equipped with an Intel\textsuperscript{\textregistered} Core\textsuperscript{\texttrademark} i7-8750~CPU at \SI{2.2}{GHz} with $6$~cores, and a NVIDIA\textsuperscript{\textregistered} GeForce GTX \textsuperscript{\textregistered} 1050 Ti GPU with \SI{4}{GB} of memory. 

\subsection{Interpretability}
\label{subsec:interpretability-methods}

The HIFIS-RNN-MLP model was a neural network and therefore it was not inherently interpretable. This is why neural networks are called "black box" models. To increase model transparency, the Local Interpretable Model-Agnostic Explanations (LIME) method was applied~\cite{Ribeiro2016}. LIME was created to explain predictions made by any "black box" models. The basic principle of LIME is the assumption that nonlinear models may be approximated by linear models at a small scale. LIME slightly perturbs the feature values of an example, creating a set of similar examples within its neighbourhood. An exponential kernel is used to define the neighbourhood. Any inherently interpretable model trained on the black box's predictions of the generated neighbourhood of examples, may then be used to approximate the black box model's functionality within that neighbourhood. The local model's parameters are taken to represent the relative importance of the original example's feature values to the original model's prediction.

Many parameters of LIME were evaluated in pursuit of transparent and human-friendly explanations. Ridge regression was chosen as the inherently interpretable local model to explain the HIFIS-RNN-MLP predictions using LIME. To decrease the feature space of the local surrogate model, the numerical features were discretized into $4$ bins. Next, the choice of kernel width is crucial because it defines the size of the neighbourhood generated around an example, which greatly influences the locality and stability of LIME explanations. Stability is a property of explanations that refers to how alike explanations for similar examples are~\cite{Molnar2019}. Since we prioritized stability, we endeavoured to produce explanations that should minimally differ when produced for the same client. Too small a kernel width increased locality of explanations and compromised their stability. Whereas, increasing the kernel width too much can cause the local model to approach a global surrogate, which is counter to the goal of local explanations in the first place~\cite{Gruber2019}. After investigating different values for the kernel width, we chose to use the default choice in the author's implementation of LIME, which is $0.75 \cdot \sqrt{|\vec{x}|}$. The default value produced local and reasonably stable explanations. To further enhance explanation stability, we increased the sample size, which refers to the number of slightly perturbed examples generated and used to fit the local model. According to an experimental analysis of LIME by Molnar~\etal, "the sample size was a strictly monotonous benefactor for explanation stability and thus should not be reduced"~\cite{Gruber2019}. Accordingly, we set the sample size to \num{40000}, which represented the ceiling of computational overhead we were willing to accept given production deployment requirements. With the aforementioned values for LIME parameters, each explanation took approximately \SI{8.4}{seconds} to compute using our hardware. Overall, the application of LIME with our chosen set of parameters yielded sufficiently local and stable explanations.

\section{Results}
\label{sec:results}

To assess the performance of the HIFIS-RNN-MLP model, various metrics were considered, including recall, precision, F1-score, AUC, and our weighted F1 loss. When conducting training experiments, it was necessary to evaluate and fine-tune towards models that were consistent with Homeless Prevention's goals. Consider that a false negative corresponds to a scenario in which a client becomes chronically homeless within the next 6 months, despite the model's prediction that they were not at risk. Missing at risk individuals is highly undesirable. Alternatively, a false positive case corresponds to a situation in which a client is predicted by the model to be at high risk of chronic homelessness in the next $6$ months, but does not end up becoming chronically homeless in the future. Given a choice, the latter scenario is preferred by Homeless Prevention. The cost of a false negative is much higher than the cost of a false positive since effective preventive resources can save significant costs incurred by would-be long-term shelter users~\cite{Culhane2011}. These emotional and financial costs savings can be realized to the fullest extent if our model greatly reduces the number of clients who are falsely misclassified as unlikely to be chronically homeless in the future. Models were therefore selected using recall, precision and F1-score as the primary evaluation criteria, with a preference toward recall to minimize false negatives.

\subsection{Model Performance}
\label{subsec:model-performance}

\begin{figure}[h!]
    \centering
    \includegraphics[width=\linewidth]{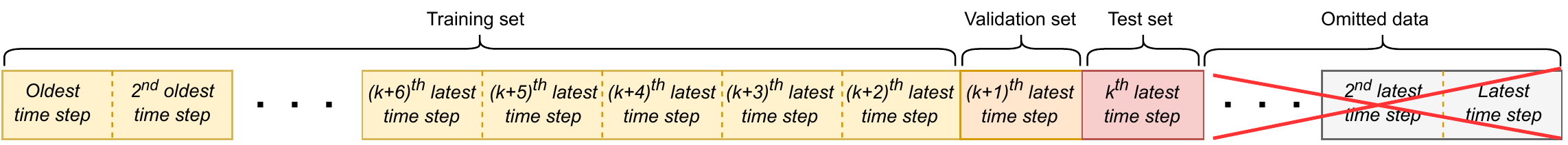}
    \caption{Dataset partitioning for the $k\textsuperscript{th}$ fold in the nested cross validation. Each block represents all existing client examples for a particular time step.}
    \label{fig:nested-cv} 
\end{figure}

The model was evaluated based on average performance on held out data via cross validation. Since the model addresses a forecasting problem, traditional partitions for validation folds do not apply. A form of nested cross validation was implemented that draws inspiration from rolling-origin evaluation described by Tashman~\etal~\cite{Tashman2000} and rolling-origin-recalibration evaluation discussed by Bergmeir~\etal~\cite{Bergmeir2012}. For each fold, the dataset was partitioned by assigning records from the second-most and most recent time steps to the validation and test respectively, with all earlier records comprising the training set. In the first fold, the entire dataset was partitioned. For the $k\textsuperscript{th}$ fold, records from the $k - 1$ most recent time steps were omitted prior to partitioning. Refer to Figure \ref{fig:nested-cv} for a portrayal of dataset partitioning for the $k\textsuperscript{th}$ fold. The model was trained on $10$ folds defined by this nested cross validation method.

To endorse the decision of framing the problem using both dynamic features and static features, we trained a MLP model that did not take into account dynamic features. Only running totals of service features were considered. In this static data modality, each example consisted of the static features for client, calculated as per their records in the HIFIS database available at the time of writing. The static dataset was indexed by \textit{ClientID}, and could therefore be evaluated using k-fold cross validation by defining test sets composed of held out client records. Test set results from a 10-fold cross validation are reported in Table \ref{tab:model-compare}. As demonstrated by the results, recall was comparable for both models, but the HIFIS-RNN-MLP model achieved considerably higher precision than the MLP model using the static dataset.

\begin{table}[h]
  \centering
  \begin{tabular}{c c c c c c c}
    \toprule
    \textit{Features}                  & \textit{Model}      & \multicolumn{5}{c}{\textit{Mean metric value [standard deviation]} $\times 10^2$} \\
                                       &                     & Recall & Precision & F1-score & AUC & Accuracy \\
    \midrule
    \multirow{3}{*}{Dynamic~\&~static} & \textbf{HIFIS-RNN-MLP}       & \textbf{\num{92.1} [\num{1.7}]} & \textbf{\num{65.1} [\num{3.0}]} & \textbf{\num{76.3} [\num{2.0}]} & \textbf{\num{97.6} [\num{0.7}]} & \textbf{\num{97.1} [\num{0.2}]}  \\
                                       & Logistic regression & \num{93.2} [\num{1.8}] & \num{61.7} [\num{1.9}] & \num{74.2} [\num{1.8}] & \num{98.9} [\num{0.3}] & \num{96.7} [\num{0.2}]   \\
                                       & Random forest       & \num{74.0} [\num{5.1}] & \num{87.2} [\num{1.1}] & \num{80.0} [\num{3.1}] & \num{99.1} [\num{0.3}] & \num{98.1} [\num{0.1}]   \\
    
    \midrule
    \multirow{3}{*}{Static}            & MLP                 & \num{89.7} [\num{5.0}] & \num{36.3} [\num{06.5}] & \num{51.3} [\num{6.3}] & \num{96.5} [\num{1.3}] & \num{92.2} [\num{1.0}]   \\
                                       & Logistic regression & \num{80.6} [\num{7.0}] & \num{38.9} [\num{4.4}] & \num{52.3} [\num{5.1}] & \num{95.0} [\num{2.4}] & \num{93.2} [\num{0.8}]   \\
                                       & Random forest       & \num{17.0} [\num{9.5}] & \num{60.8} [\num{17.4}] & \num{25.8} [\num{12.8}] & \num{95.6} [\num{1.8}] & \num{95.7} [\num{0.9}]   \\
    \bottomrule
  \end{tabular}
  \vspace{8pt}
  \caption{Holdout performance for cross validation of various models and data modalities}
  \vspace{-10pt}
  \label{tab:model-compare}
\end{table}

To further illustrate the utility of the HIFIS-RNN-MLP model, its performance was compared to classical learning algorithms, logistic regression and a $100$-tree random forest. These benchmark models were trained on both the dynamic time series dataset and the static dataset and results of cross validation of all models considered is reported in Table \ref{tab:model-compare}. Class weighting was employed in logistic regression, random forest, and the MLP, to imbue training with our goal of accurately identifying positives. However, HIFIS-RNN-MLP trained using the weighted F1 loss was able to achieve a balance of recall and precision closest to Homeless Prevention's goals.

To demonstrate the utility of the custom F1 loss function, in Table~\ref{tab:loss-comparison} we display the results of cross validation performance of the HIFIS-RNN-MLP model using various formulations of the loss function. As well as compared against binary cross entropy (BCE) as a loss function. Since the dataset was unbalanced, class weighting was investigated as a means to force more attention to be paid to the minority positive class examples. When conducting training experiments with weighted BCE loss, a penalty was applied based on the fraction of examples with positive ground truth (i.e. $6.56\%$). The results in Table~\ref{tab:loss-comparison} indicate that the custom F1 loss function is slightly more effective than class-weighted BCE at achieving the most desirable precision-recall balance.

\begin{table}[h]
  \centering
  \begin{tabular}{c c c c c c}
    \toprule
    \textit{Loss function}                  & \multicolumn{5}{c}{\textit{Mean metric value [standard deviation]} $\times 10^2$} \\
                                                            & Recall & Precision & F1-score & AUC & Accuracy \\
    \midrule
    \textbf{Weighted F1 loss}          & \textbf{\num{92.1} [\num{1.7}]} & \textbf{\num{65.1} [\num{3.0}]} & \textbf{\num{76.3} [\num{2.0}]} & \textbf{\num{97.6} [\num{0.7}]} & \textbf{\num{97.1} [\num{0.2}]}  \\
    BCE with class weighting           & \num{93.5} [\num{001.4}] & \num{63.8} [\num{03.8}] & \num{75.8} [\num{2.7}] & \num{99.1} [\num{0.2}] & \num{97.0} [\num{0.3}]   \\
    BCE           & \num{77.3} [\num{5.2}] & \num{84.6} [\num{2.7}] & \num{80.7} [\num{03.0}] & \num{99.1} [\num{0.2}] & \num{98.1} [\num{0.2}]   \\
    \bottomrule
  \end{tabular}
  \vspace{8pt}
  \caption{Holdout set performance for cross validation of HIFIS-RNN-MLP with varying loss functions.}
  \vspace{-10pt}
  \label{tab:loss-comparison}
\end{table}

\subsection{Interpretability}
\label{subsec:interpretability-results}

\begin{figure}[h!]
    \centering
    \begin{subfigure}{0.495\textwidth}
        \centering
        \includegraphics[width=\textwidth]{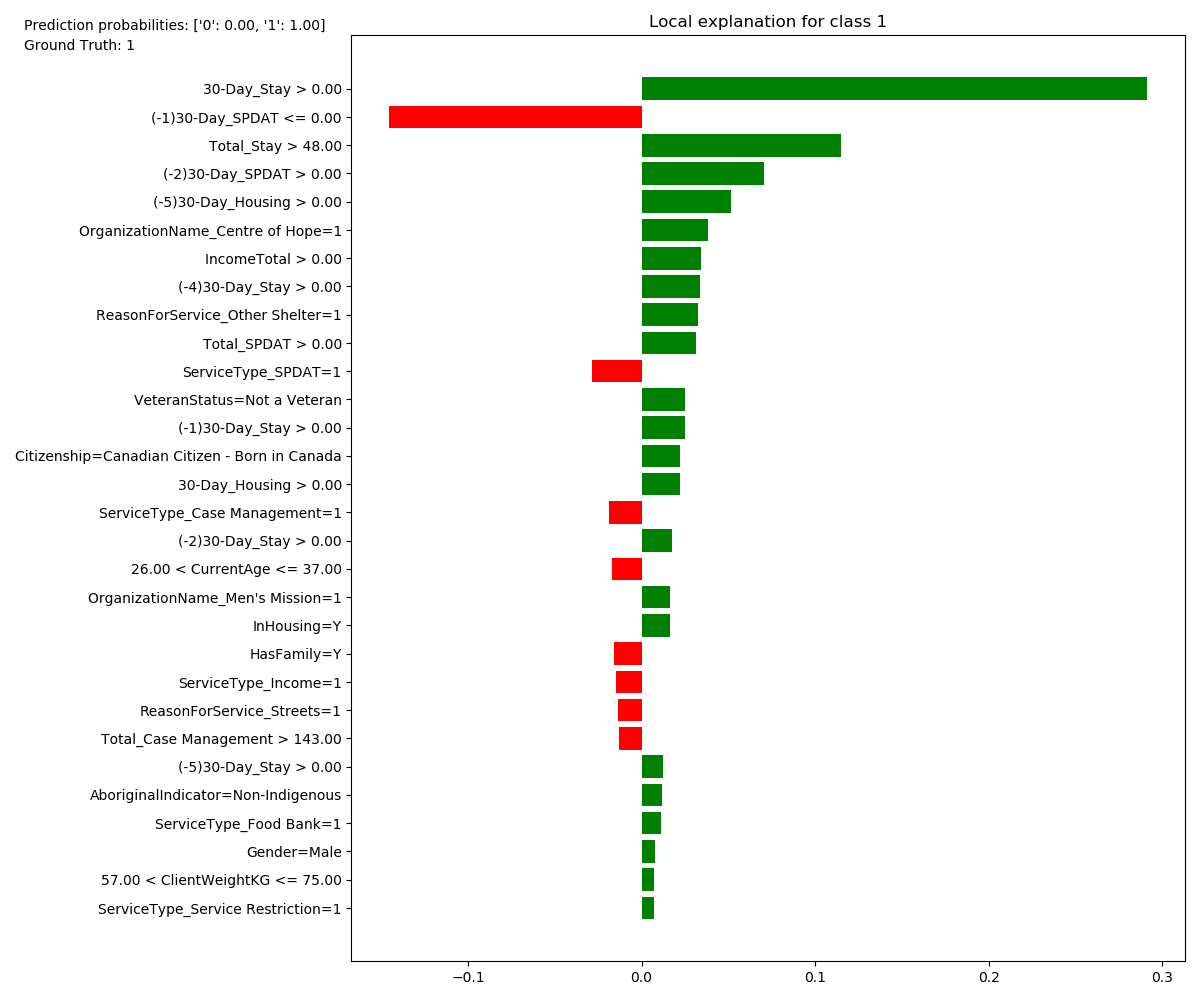}
        \label{fig:explanation-1}
    \end{subfigure}
    \hfill
    \begin{subfigure}{0.495\textwidth}
        \centering
        \includegraphics[width=\textwidth]{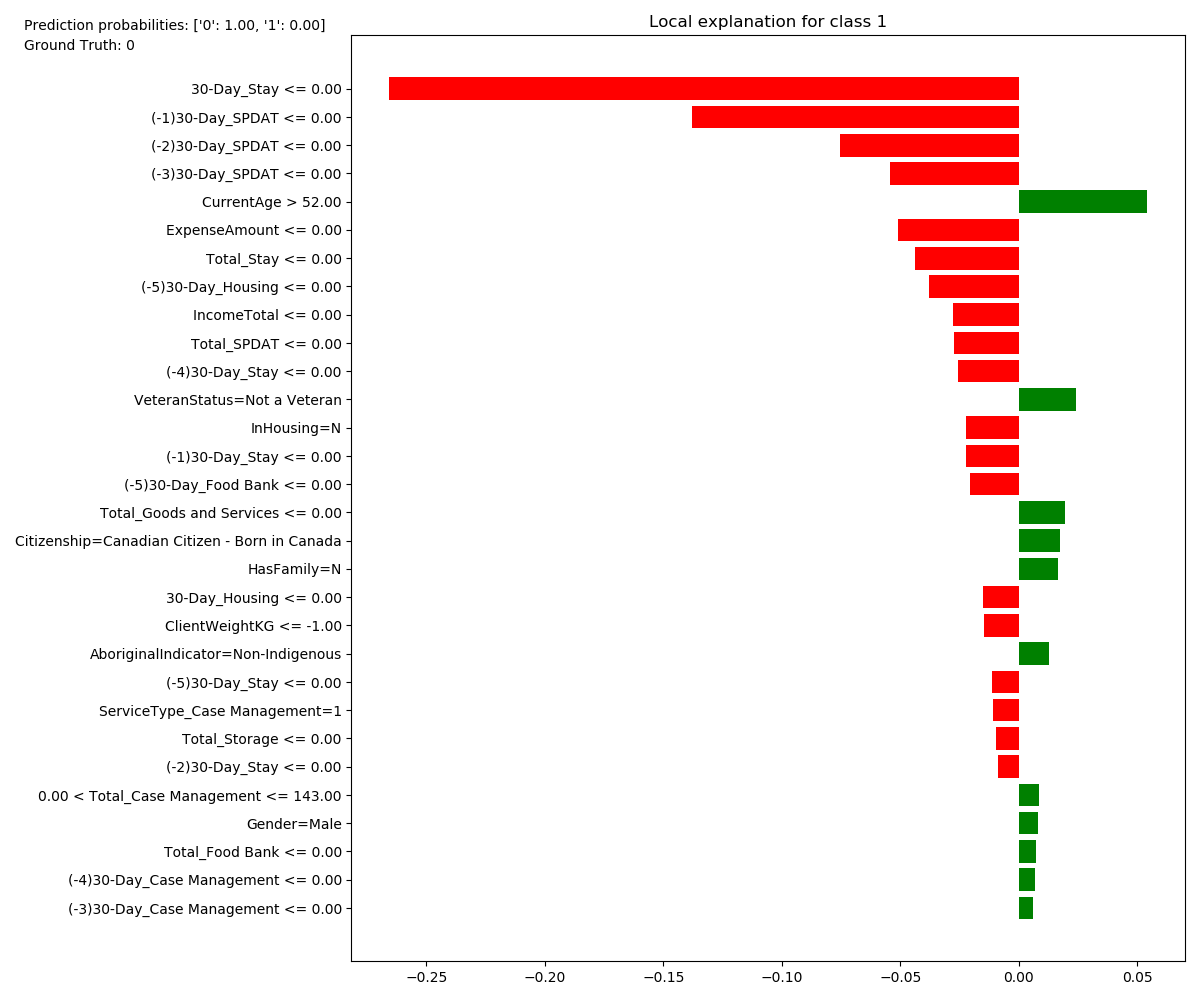}
        \label{fig:explanation-4}
    \end{subfigure}
    \caption{A selection of LIME explanations for client examples in the test set. Explanations consist of a list of feature values with weight corresponding to the bars on the graphs. Green and red bars indicate contribution towards and against a prediction of chronic homelessness respectively.}
    \label{fig:lime-explanations}
\end{figure}

The application of LIME to predictions made by the HIFIS-RNN-MLP model resulted in explanations that were not only consistent with predictors of chronic homelessness from the literature, but provided additional insight into the chronically homeless population specific to London. Each explanation consists of a series of paired feature values and weights, listed in descending order by weight. The feature values with the highest weight magnitudes may be considered as the client's attributes that most contributed toward the model's prediction of a positive ground truth. A selection of examples of client explanations are shown in Figure \ref{fig:lime-explanations}. The local explanations for each client served three main purposes in the context of model development. First, they helped ensure that unintended bias was not present in the model's decisions. Explanations were examined to ensure that predictions were not contingent on single demographic features. Second, explanations aided in iterative feature engineering. Some categorical features take on several possible values, and thus constitute large fractions of a preprocessed example. Any categorical features that never appeared in explanations were subsequently appended to the list of features to exclude prior to training (e.g. medications). Finally, collaborative analysis of explanations with domain experts at Homeless Prevention helped validate that the model was not fixating on bizarre or unrealistic correlations.

\begin{figure}[h!]
    \centering
    \includegraphics[width=\linewidth]{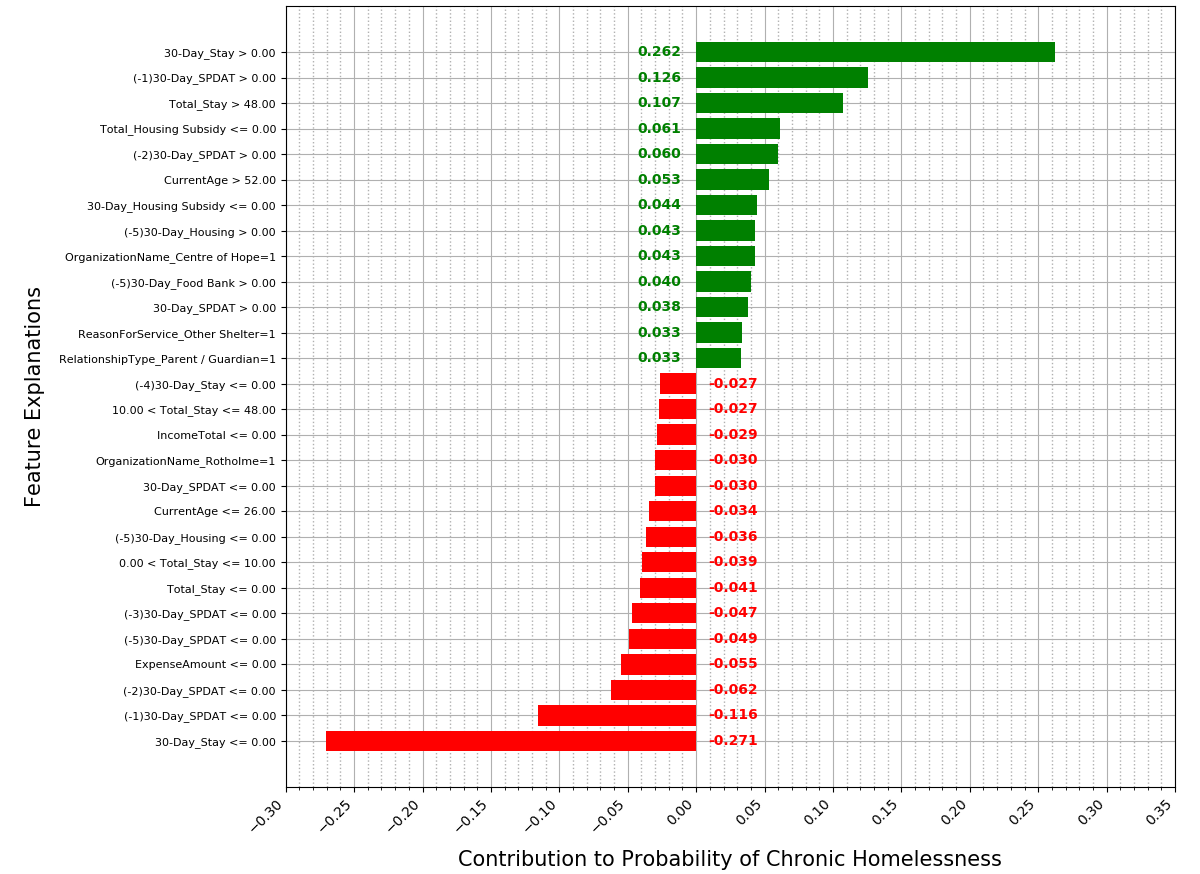}
    \caption{Results of the LIME submodular pick procedure. This graph communicates an approximation of model functionality. Each bar corresponds to the weight of a feature value or range. Green and red bars indicate contribution toward and against prediction of chronic homelessness. The magnitude of a bar indicates its relative influence in the model's decision.}
    \label{fig:submodular-pick} 
\end{figure}

Although individual explanations are highly informative in and of themselves, their value in understanding the model as a whole is limited. The authors of LIME proposed a method called \textit{submodular pick}, which was designed to provide a holistic understanding of the model by combining a series of explanations that maximally covers the model's input space~\cite{Ribeiro2016}. The submodular pick algorithm, after computing a tuneable number of explanations, performs a greedy pick of explanations that maximizes the representation of the input feature space~\cite{Ribeiro2016}. The result is a list of feature values paired with weights. An explanation's weights approximate the relative importance of feature values to the model's decision, independent of a specific example. Submodular pick was executed with $20\%$ of the combined training and validation sets as input. The resultant feature values and weights constitute the global model explanation depicted in Figure \ref{fig:submodular-pick}. Positive and negative weights correspond to contribution towards and against a prediction of chronic homelessness respectively. The most conspicuous outcome in Figure \ref{fig:submodular-pick} is the importance of the number of stays in the most recent $30$-day~time~step (i.e. \textit{"30-Day\_Stay"}). Similarly, total shelter stays were a very influential feature (i.e. \textit{"Total\_Stay"}). These findings corroborate Shinn~\etal's result that previous shelter stays were the strongest predictive feature for familial homelessness~\cite{Shinn2013}. It also appears that the administration of a SPDAT screening questionnaire $1$~time~step ago is highly predictive of chronic homelessness (i.e. \textit{"(-1)30-Day\_SPDAT"}), perhaps commending London case workers' ability to identify high-risk clients weeks prior to transitioning to chronic homelessness as defined here. Also of remarkable magnitude is a client's aggregate days of receipt of housing subsidies (i.e. \textit{"Total\_Housing\_Subsidy"}). According to the global explanation, lack of receipt of housing subsidies steers the the model towards predicting chronic homelessness, which is in keeping with Byrne~\etal's finding of negative association between chronic homelessness and subsidized housing~\cite{Byrne2014}. Again corroborating the work of Shinn~\etal, advanced age appears to be a predictive factor~\cite{Shinn2013}. As reported in Figure \ref{fig:submodular-pick}, being in the highest age bin \textit{"CurrentAge~>~52.00"}) increases a client's risk of future chronic homelessness; whereas, belonging to the lowest age bin \textit{"CurrentAge~<=~26.00"}) seems protective.

\section{Discussion}
\label{sec:discussion}

This project is among the first to apply an artificial neural network to model chronic homelessness. Moreover, it succeeded in illuminating the insights learned by the neural network, which are traditionally viewed as "black box" models. Performance metrics achieved by the HIFIS-RNN-MLP model exceeded Homeless Prevention's expectations. Among the work most similar to ours is the research by Toros~\etal that developed separate models to predict persistent homelessness among adults who exited the job market recently and young adults receiving public benefits~\cite{Toros2019}. Their criterion for "persistently homeless" is having experienced more than $1$ period of homelessness (defined as having no address) within $3$ years. Despite not being equivalent to the "chronically homeless" state designated in this study, we consider their problem similar enough to warrant comparison. Toros~\etal trained $2$ models that attained a holdout AUC of $0.89$ and $0.88$~\cite{Toros2019}. Their employment and young adult models achieved recall of $0.308$ and $0.351$ at classification thresholds of $0.528$ and $0.471$ respectively. Nested cross validation of the HIFIS-RNN-MLP model yielded an AUC of $0.976$ and a recall of $0.921$ at a classification threshold of $0.5$.

The predictive capability of the HIFIS-RNN-MLP model introduces the possibility of early identification of clients who are at risk of chronic homelessness. In other jurisdictions, preventive strategies such as housing subsidies, diversionary efforts and community-based services are shown to be effective~\cite{Shinn}. In their publication presenting a new framework for homelessness prevention in Canada, Gaetz and Dej present multiple arguments that preventative strategies are cost-effective for society~\cite{Gaetz2017}. Avoiding emotional and physical trauma for the chronically homeless population in the shelter system and conserving finite public resources makes these efforts vital. It stands to reason that an interpretable machine learning algorithm that sharpens case workers' ability to identify high risk individuals who would benefit from preventative resources would be invaluable to any municipality. Further, client predictions may be accumulated to forecast aggregate shelter demand. Aside from the immediately tangible benefits, the interpretable nature of this model may help service providers more deeply understand factors contributing to chronic homelessness in their locale.

Our work is readily replicable by other cities who use the HIFIS application. The Canadian government mandated that all municipalities have a homelessness information management system, and offered subsidized implementation of HIFIS to those lacking one~\cite{Canada2020}. Using their own HIFIS database and the open source code accompanying this paper, municipalities could apply the methods described herein. Care was taken to thoroughly document the open source repository and adhere to modular design so as to enable quick adaptation and implementation. Further the model does not require a GPU to train in an acceptable length of time which further decreases the barrier to implementation from a compute infrastructure perspective. 

A key facet of this work was the application of LIME to probe the model for explanations. Interpretability is fundamental to the ethical deployment of decision-making algorithms in the public sector. Methods such as LIME promote transparency and identification of sources of unintended bias. The Canadian government in their Directive on Automated Decision Making states that some automated decision-making systems provide explanations that justify their choices~\cite{Canada2019}. As an interpretability method, LIME fit this application space. Not only is it possible to obtain explanations on granular example-wise basis, but an overall understanding of the model's behaviour may also be realized by studying the results of the submodular pick algorithm. LIME provides insight into which specific features of a client are most relevant to the decision made by the model, enabling service providers to evaluate a client's prediction in the context of their history. Although previous studies applied inherently interpretable machine learning methods to homelessness prevention (e.g. logistic regression, decision trees), the HIFIS-RNN-MLP model demonstrated performance metric superiority and was interpretable through the utilization of LIME.

Despite the encouraging results of this study, it was not free of limitations. First, the advent of HIFIS in London was relatively recent; as a result, we only had approximately \SI{4}{years} of records to access. With a time step of \SI{30}{days}, our dataset contained \num{115515} records for \num{6521} clients. In contrast, Toros~\etal's investigation amassed records for recipients of public benefits, children and family public service usage, and homelessness information management system data spanning $10$, $8$, and \SI{5}{years} respectively~\cite{Toros2019}. Additionally, the problem statement of this study limits the model to predicting transition into only the chronic state of homelessness. By definition, episodic and transitional states of homelessness are left out. Finally, our study's definition of chronic homelessness fails to capture the variety of states that chronically homeless people may be in when staying outside the shelter system. Sleeping rough, couch surfing, and stays in healthcare institutions were not accounted for in the calculation of the ground truth. Hence, the ground truth may fail to capture individuals who are homeless for the majority of the year, but stay in a shelter only for a minority of the year.

A possible enhancement to the methods could be the exploration of interpretability methods other than LIME, such as partial dependence plots~\cite{Friedman2001} or SHAP~\cite{Lundberg2017}. Due to LIME meeting the interpretable requirements of service providers, no others were investigated. Further study of the model should include its evaluation in deployment. To this end, a randomized control trial could be conducted to quantify any additional preventative resources deployed to clients as a result of the model's predictions, compared to a control group. Next, a direct comparison of the HIFIS-RNN-MLP model and the SPDAT would support the assertion that this model is a feasible decision support algorithm. In addition to comparing performance, a comparison of the SPDAT's highly weighted features to those highlighted in a LIME submodular pick of our model would be of great interest. To conduct a true comparison, all client features relating to the SPDAT could be excluded from our model. As the SPDAT's use is widespread, this proposed analysis may enhance trust in our study's methods outside of London, Canada. Finally, future investigations of modelling the homeless population in London could include locations occupied by homeless individuals other than shelters, and could incorporate episodic and transitional states of homelessness in the formulation of the ground truth.

\section{Conclusion}
\label{sec:conclusion}
In this project, a machine learning model was trained to effectively predict chronic homelessness among individuals receiving services in London, Canada. Dubbed HIFIS-RNN-MLP, this model connected RNN and MLP architectures to process dynamic and static features. The trained model achieved a mean recall of $0.921$ and precision of $0.651$ across the holdout sets of a $10$-fold nested cross validation. Application of the LIME interpretability algorithm yielded local explanations that met the requirements of stakeholders. Execution of the submodular pick algorithm produced a global LIME explanation that approximates rules that the model learned from the combinations of input features. Our methods are reproducible and freely accessible. It is our hope that other municipalities may derive benefit from this work.

\section{Acknowledgements}
\label{sec:acknowledgements}

We wish to express gratitude for the support and expertise of the following individuals throughout this project: Mat Daley, Vala Gylfadottir, Craig Cooper, Trevor Fowler and Bryan Knight.

\bibliographystyle{unsrt}  
\small{
\bibliography{references}  

\begin{thebibliography}{10}

\bibitem{Gaetz2016}
Stephen Gaetz, Erin Dej, Tim Richter, and Melanie Redman.
\newblock {\em {The State of Homelessness in Canada 2016}}.
\newblock 2016.

\bibitem{Canada2020}
{Government of Canada}.
\newblock {Reaching Home: Canada's Homelessness Strategy Directives}, 2020.

\bibitem{Shinn}
Marybeth Shinn and Rebecca Cohen.
\newblock {Homelessness Prevention: A Review of the Literature}.
\newblock Technical report, 2019.

\bibitem{SPDAT}
{Vulnerability Index - Service Prioritization Decision Assistance Tool
  Prescreen Triage Tool for Single Adults Welcome to the SPDAT Line of Products
  VI-SPDAT Series}.
\newblock Technical report, OrgCode Consulting Inc., 2015.

\bibitem{Brown2018}
Molly Brown, Camilla Cummings, Jennifer Lyons, Andr{\'{e}}s Carri{\'{o}}n, and
  Dennis~P. Watson.
\newblock {Reliability and validity of the Vulnerability Index-Service
  Prioritization Decision Assistance Tool (VI-SPDAT) in real-world
  implementation}.
\newblock {\em Journal of Social Distress and the Homeless}, 27(2):110--117,
  2018.

\bibitem{Shinn2013}
Marybeth Shinn, Andrew~L. Greer, Jay Bainbridge, Jonathan Kwon, and Sara
  Zuiderveen.
\newblock {Efficient targeting of homelessness prevention services for
  families}.
\newblock {\em American Journal of Public Health}, 103(SUPPL. 2):324--330,
  2013.

\bibitem{Hong2018}
Boyeong Hong, Awais Malik, Jack Lundquist, Ira Bellach, and Constantine~E.
  Kontokosta.
\newblock {Applications of Machine Learning Methods to Predict Readmission and
  Length-of-Stay for Homeless Families: The Case of Win Shelters in New York
  City}.
\newblock {\em Journal of Technology in Human Services}, 36(1):89--104, 2018.

\bibitem{Greer2016}
Andrew Greer, Marybeth Shinn, Jonathan Kwon, and Sara Zuiderveen.
\newblock {Targeting services to individuals most likely to enter shelter:
  Evaluating the efficiency of homelessness prevention}.
\newblock {\em Social Service Review}, 90(1):130--155, 2016.

\bibitem{Flaming2011}
Daniel Flaming, Patrick Burns, Gerald Sumner, Manuel~H. Moreno, and Halil
  Toros.
\newblock {Crisis Indicator: Triage Tool for Identifying Homeless Adults in
  Crisis}.
\newblock Technical report, Economic Roundtable, 2011.

\bibitem{Toros2017}
Halil Toros and Daniel Flaming.
\newblock {Prioritizing Which Homeless People Get Housing Using Predictive
  Algorithms}.
\newblock {\em SSRN Electronic Journal}, pages 1--32, 2017.

\bibitem{Toros2019}
Halil Toros, Daniel Flaming, and Patrick Burns.
\newblock {Early Intervention to Prevent Persistent Homelessness}.
\newblock Technical Report March, Economic Roundtable, 2019.

\bibitem{VonWachter2019}
Till {Von Wachter}, Marianne Bertrand, Harold Pollack, Janey Rountree, and
  Brian Blackwell.
\newblock {Predicting and Preventing Homelessness in Los Angeles}.
\newblock Technical Report September, The California Policy Lab, University of
  Chicago Poverty Lab, 2019.

\bibitem{Chan2017}
Hau Chan, Eric Rice, Phebe Vayanos, Milind Tambe, and Matthew Morton.
\newblock {Evidence from the past: AI decision AIDS to improve housing systems
  for homeless youth}.
\newblock {\em AAAI Fall Symposium - Technical Report}, FS-17-01 -:149--157,
  2017.

\bibitem{Fisher2020}
Andrew Fisher, Vijay Mago, and Eric Latimer.
\newblock {Simulating the evolution of homeless populations in canada using
  modified deep Q-Learning (MDQL) and modified neural fitted Q-Iteration (MNFQ)
  algorithms}.
\newblock {\em IEEE Access}, 8:92954--92968, 2020.

\bibitem{LeCun1998}
Yann LeCun, Leon Bottou, Genevieve~B Orr, and Klaus-Robert M{\"{u}}ller.
\newblock {Efficient BackProp}.
\newblock In Gerhard Goos, Juris Hartmanis, and Jan van Leeuwen, editors, {\em
  Neural Networks Tricks of the Trade}, chapter~1, pages 9--50. Springer-Verlag
  Berlin Heidelberg, 1998.

\bibitem{Bergmeir2012}
Christoph Bergmeir and Jos{\'{e}}~M. Ben{\'{i}}tez.
\newblock {On the use of cross-validation for time series predictor
  evaluation}.
\newblock {\em Information Sciences}, 191:192--213, 2012.

\bibitem{Hsu2019}
Te~Cheng Hsu, Shing~Tzuo Liou, Yun~Ping Wang, Yung~Shun Huang, and Che-Lin.
\newblock {Enhanced Recurrent Neural Network for Combining Static and Dynamic
  Features for Credit Card Default Prediction}.
\newblock {\em ICASSP, IEEE International Conference on Acoustics, Speech and
  Signal Processing - Proceedings}, 2019-May:1572--1576, 2019.

\bibitem{Karpathy2019}
{Andrej Karpathy}.
\newblock {A Recipe for Training Neural Networks}, 2019.

\bibitem{Srivastava2014}
Nitish Srivastava, Geoffrey Hinton, Alex Krizhevsky, Ilya Sutskever, and Ruslan
  Salakhutdinov.
\newblock {Dropout: A Simple Way to Prevent Neural Networks from Overfitting}.
\newblock {\em Journal of Machine Learning Research}, 15(56):1929--1958, 2014.

\bibitem{Prechelt1998}
Lutz Prechelt.
\newblock {Early Stopping - But When?}
\newblock In Gerhard Goos, Juris Hartmanis, and Jan van Leeuwen, editors, {\em
  Neural Networks Tricks of the Trade}, chapter~2, pages 55--69.
  Springer-Verlag Berlin Heidelberg, 1998.

\bibitem{Kingma2014}
Diederik~P. Kingma and Jimmy Ba.
\newblock Adam: {A} method for stochastic optimization.
\newblock In Yoshua Bengio and Yann LeCun, editors, {\em 3rd International
  Conference on Learning Representations, {ICLR} 2015, San Diego, CA, USA, May
  7-9, 2015, Conference Track Proceedings}, 2015.

\bibitem{Bergstra2012}
James Bergstra and Yoshua Bengio.
\newblock {Random search for hyper-parameter optimization}.
\newblock {\em Journal of Machine Learning Research}, 13(10):281--305, 2012.

\bibitem{Ribeiro2016}
Marco~Tulio Ribeiro, Sameer Singh, and Carlos Guestrin.
\newblock {"Why should i trust you?" Explaining the predictions of any
  classifier}.
\newblock In {\em Proceedings of the ACM SIGKDD International Conference on
  Knowledge Discovery and Data Mining}, volume 13-17-Augu, pages 1135--1144,
  2016.

\bibitem{Molnar2019}
Christoph Molnar.
\newblock {\em Interpretable Machine Learning}.
\newblock 2019.
\newblock \url{https://christophm.github.io/interpretable-ml-book/}.

\bibitem{Gruber2019}
Christoph Molnar, Sebastian Gruber, and Philipp Kopper.
\newblock {\em {Limitations of Interpretable Machine Learning Methods}}.
\newblock 2019.

\bibitem{Culhane2011}
Dennis~P. Culhane, Jung~Min Park, and Stephen Metraux.
\newblock {The Patterns and Costs of Services Use Among Homeless Families}.
\newblock {\em Journal of Community Psychology}, 39(7):815--825, 2011.

\bibitem{Tashman2000}
Leonard~J. Tashman.
\newblock {Out-of-sample tests of forecasting accuracy: An analysis and
  review}.
\newblock {\em International Journal of Forecasting}, 16(4):437--450, 2000.

\bibitem{Byrne2014}
Thomas Byrne, Jamison~D. Fargo, Ann~Elizabeth Montgomery, Ellen Munley, and
  Dennis~P. Culhane.
\newblock {The relationship between community investment in permanent
  supportive housing and chronic homelessness}.
\newblock {\em Social Service Review}, 88(2):234--263, 2014.

\bibitem{Gaetz2017}
Stephen Gaetz and Erin Dej.
\newblock {\em {A New Direction: A Framework for Homelessness Prevention}}.
\newblock Canadian Observatory on Homelessness Press, Toronto, 2017.

\bibitem{Canada2019}
{Government of Canada}.
\newblock {Directive on Automated Decision-Making}, 2019.

\bibitem{Friedman2001}
Jerome Friedman.
\newblock {Greedy Function Approximation: A Gradient Boosting Machine}.
\newblock {\em The Annals of Statistics}, 29(5):1189--1232, 2001.

\bibitem{Lundberg2017}
Scott~M Lundberg and Su-In Lee.
\newblock A unified approach to interpreting model predictions.
\newblock In I.~Guyon, U.~V. Luxburg, S.~Bengio, H.~Wallach, R.~Fergus,
  S.~Vishwanathan, and R.~Garnett, editors, {\em Advances in Neural Information
  Processing Systems 30}, pages 4765--4774. Curran Associates, Inc., 2017.

\end{thebibliography}
}

\newpage    

\appendix
\section*{Appendix}

\section{Feature Descriptions}
\label{apx:feature_descriptions}

\normalsize{
Categorizations and descriptions for features of a preprocessed client example. The dataset of preprocessed examples is indexed by \textit{ClientID} and \textit{Date}. Dynamic features exist for the current $30$-day time step (TS) and for each previous time step~$\textbf{t} \in [1,T_x - 1]$.
}

\begin{table}[h!]
 \captionsetup{labelformat=empty}
  \centering
  \tiny{
  \begin{tabular}{l l l l}
    \toprule
    Temporality    &    Feature type   & Feature name        & Description \\
    \midrule
    \multirow{14}{*}{Static}
    & \multirow{14}{*}{Numerical} & CurrentAge    & Client's age (in years)   \\
      &  & ClientWeightKG    & Weight (in kilograms)   \\
      &  & ExpenseAmount     & Total routine expenses (in Canadian dollars)   \\
      &  & TotalScore    & Most recent SPDAT score, in range $[-1, 12]$. $-1$ indicates no SPDAT.  \\
      &  & Total\_Stay    & Total \# stays in a shelter \\
      &  & Total\_Case Management & Total \# days of case management \\
      &  & Total\_Housing & Total \# days in supportive housing received \\
      &  & Total\_Housing Subsidy & Total \# days receiving housing subsidy \\
      &  & Total\_Storage & Total \# days of storage service received \\
      &  & Total\_Reservations & Total \# shelter bed reservations \\
      &  & Total\_Turnaways   & Total \# times an individual was refused service at a shelter \\
      &  & Total\_Food Bank   & Total \# shelter meals \\
      &  & Total\_Goods and Services  & Total \# goods and services records \\
      &  & Total\_SPDAT   & Total \# times client has taken the SPDAT \\
    \midrule
    \multirow{6}{*}{Static} & \multirow{6}{*}{Single-valued categorical} & Gender    & Client's gender \\
    &    & AboriginalIndicator    & Indigenous status of client (or lack thereof)   \\
    &    & Citizenship   & Canadian citizenship status   \\
    &    & VeteranStatus & Type of veteran   \\
    &    & InHousing & Whether client is currently housed    \\
    &    & ExpenseFrequency  & How often routine expenses occur   \\
    &    & HasFamily & Whether client has a family on record  \\
    \midrule
    \multirow{18}{*}{Static} & \multirow{18}{*}{Multi-valued categorical} & ServiceType    & Type(s) of services client has received \\
    &    & OrganizationName    & Name(s) of organizations that client has received services from   \\
    &    & ReasonForService   & Reason(s) client has received service(s)   \\
    &    & IncomeType & Type(s) of income client receives   \\
    &    & ExpenseType & Type(s) of client's expense(s)    \\
    &    & IsEssentialYN  & Whether client's expense(s) are essential   \\
    &    & Reason & Reason(s) for past service restrictions \\
    &    & HealthIssue   & Medical conditions on client's record  \\
    &    & DiagnosedYN   & Whether client has been diagnosed for any of their medical condition(s) on record  \\
    &    & SelfReportedYN    & Whether client self-reported any of their medical condition(s) on record  \\
    &    & SelfReportedYN    & Whether any of client's medical condition(s) on record are suspected, but not diagnosed   \\
    &    & ContributingFactor    & Factor(s) contributing to client's situation   \\
    &    & LifeEvent & Significant event(s) in client's life \\
    &    & PreScreenPeriod   & Periods at which client has been screened via SPDAT   \\
    &    & BehavioralFactor  & Dangerous behaviour(s) client has exhibited \\
    &    & Severity  & Severity of behavioral factor(s)  \\
    &    & RelationshipType  & Client's family role(s)   \\
    &    & EducationLevel    & Highest reported education level(s)   \\
    \midrule
    \multirow{20}{*}{Dynamic}
    & \multirow{20}{*}{Numerical} &  30-Day\_Stay    & \# shelter stays in current TS. \\
      &  & 30-Day\_Case Management & \# days of case management in current TS. \\
      &  & 30-Day\_Housing & \# days in supportive housing in current TS. \\
      &  & 30-Day\_Housing Subsidy & \# days of housing subsidy in current TS. \\
      &  & 30-Day\_Storage & \# days of storage service in current TS \\
      &  & 30-Day\_Reservations & \# shelter bed reservations in current TS \\
      &  & 30-Day\_Turnaways   & \# shelter turnaways in current TS \\
      &  & 30-Day\_Food Bank   & \# shelter meals in current 30-day TS \\
      &  & 30-Day\_Goods and Services  & \# goods and services in current TS \\
      &  & 30-Day\_SPDAT   & \# times client took SPDAT in current TS \\
      &  & (-\textbf{t})30-Day\_Stay    & \# shelter stays in \textbf{t}\textsuperscript{th} past TS \\
      &  & (-\textbf{t})30-Day\_Case Management & \# days of case management in \textbf{t}\textsuperscript{th} past TS \\
      &  & (-\textbf{t})30-Day\_Housing & \# days in supportive housing in \textbf{t}\textsuperscript{th} past TS \\
      &  & (-\textbf{t})30-Day\_Housing Subsidy & \# days of housing subsidy in \textbf{t}\textsuperscript{th} past TS \\
      &  & (-\textbf{t})30-Day\_Storage & \# days of storage service in \textbf{t}\textsuperscript{th} past TS \\
      &  & (-\textbf{t})30-Day\_Reservations & \# shelter bed reservations in \textbf{t}\textsuperscript{th} past TS \\
      &  & (-\textbf{t})30-Day\_Turnaways   & \# shelter turnaways in \textbf{t}\textsuperscript{th} past TS \\
      &  & (-\textbf{t})30-Day\_Food Bank   & \# shelter meals in \textbf{t}\textsuperscript{th} past TS \\
      &  & (-\textbf{t})30-Day\_Goods and Services  & \# goods and services in \textbf{t}\textsuperscript{th} past TS \\
      &  & (-\textbf{t})30-Day\_SPDAT   & \# times client took SPDAT in \textbf{t}\textsuperscript{th} past TS \\
    \bottomrule
  \end{tabular}
  }
  \label{tab:feature-descriptions}
\end{table}

\end{document}